\documentclass{ws-ijmpa}
\usepackage{times}
\usepackage{overcite}

\begin{document}

\markboth{E.~Korosteleva et al.}
         {Primary Energy Measurement with EAS Cherenkov Light}

\catchline{}{}{}{}{}

\title{PRIMARY ENERGY MEASUREMENT WITH EAS CHERENKOV LIGHT:
       EXPERIMENT QUEST AND CORSIKA SIMULATION}

\author{E.~KOROSTELEVA, L.~KUZMICHEV, V.~PROSIN}
\address{Skobeltzyn Institute of Nuclear Physics, MSU,
         Moscow, Russia}

\author{B.~LUBSANDORZHIEV}
\address{Institute of Nuclear Research, RAS,
         Moscow, Russia}  

\author{EAS-TOP COLLABORATION}
\address{INFN, IFSI, Universita' Torino, Italy}

\maketitle

\pub{Received (Day Month Year)}{Revised (Day Month Year)}

\begin{abstract}
 A new primary mass independent method of energy measurement has been
 developed by exploiting: a) the joint analysis of the shower size,
 obtained by EAS-TOP, and of the EAS atmospheric Cherenkov light
 lateral distribution, obtained by the QUEST array, and b) simulations
 based on the CORSIKA/QGSJET code. The method is based on the correlation
 between the size/energy ratio and the steepness of Cherenkov light lateral
 distribution and has been compared with a ``classical'' one based on the
 Cherenkov light flux at a fixed distance ($175$~m) from the EAS core. An
 absolute energy calibration of the EAS atmospheric Cherenkov light flux
 has been obtained.
\end{abstract}


\section{Experiment QUEST and Simulations}

The QUEST experiment was developed to combine atmospheric Cherenkov
light data to the charged particle EAS-TOP measurements 
(Gran Sasso, Italy, 2000 m a.s.l.)\cite{1}. The Cherenkov light
detector was based upon five wide angle QUASAR-370 ($37$~cm diameter)
semispheric photomultipliers installed on five telescopes (average
pointing at the direction $\theta=34^\circ$, $\varphi=167^\circ$,
full field of view for $\theta\leq40^\circ$, used in analysis, is $0.41$~sr).

A new fitting function has been used to derive two main parameters of EAS 
Cherenkov light LDF for every recorded event: the light flux at core distance
of $175$~m $Q_{175}$ and the LDF steepness, defined as the ratio of the
fluxes at $100$ and $200$~m from the axis\cite{1}: $P=Q(100)/Q(200)$.

The energy measurement methods are based on parameterizations of
Cherenkov light lateral distributions obtained from simulations performed
through CORSIKA/QGSJET\cite{2,3,4} for primary protons and iron nuclei
with total energy $1, 2, 4$ and $8$~PeV, and zenith angles $\theta$
from $0^\circ$ to $39^\circ$. 

The experiment has been additionally simulated by using such parameterizations
and extracting the primary energy, the depth of EAS maximum $X_{\max}$, shower
axis direction and coordinates. The primary composition is assumed $50$\% $p$
and $50$\% Fe. All other parameters have been calculated as functions of the
main (i.e. $P=F_1\left(X_{\max},\theta,A\right)$, $N_e=F_2\left(E_0,P,A\right)$ 
and so on). All the known apparatus errors and fluctuations have been taken
into account.      

\begin{figure}[t]
 \epsfig{file=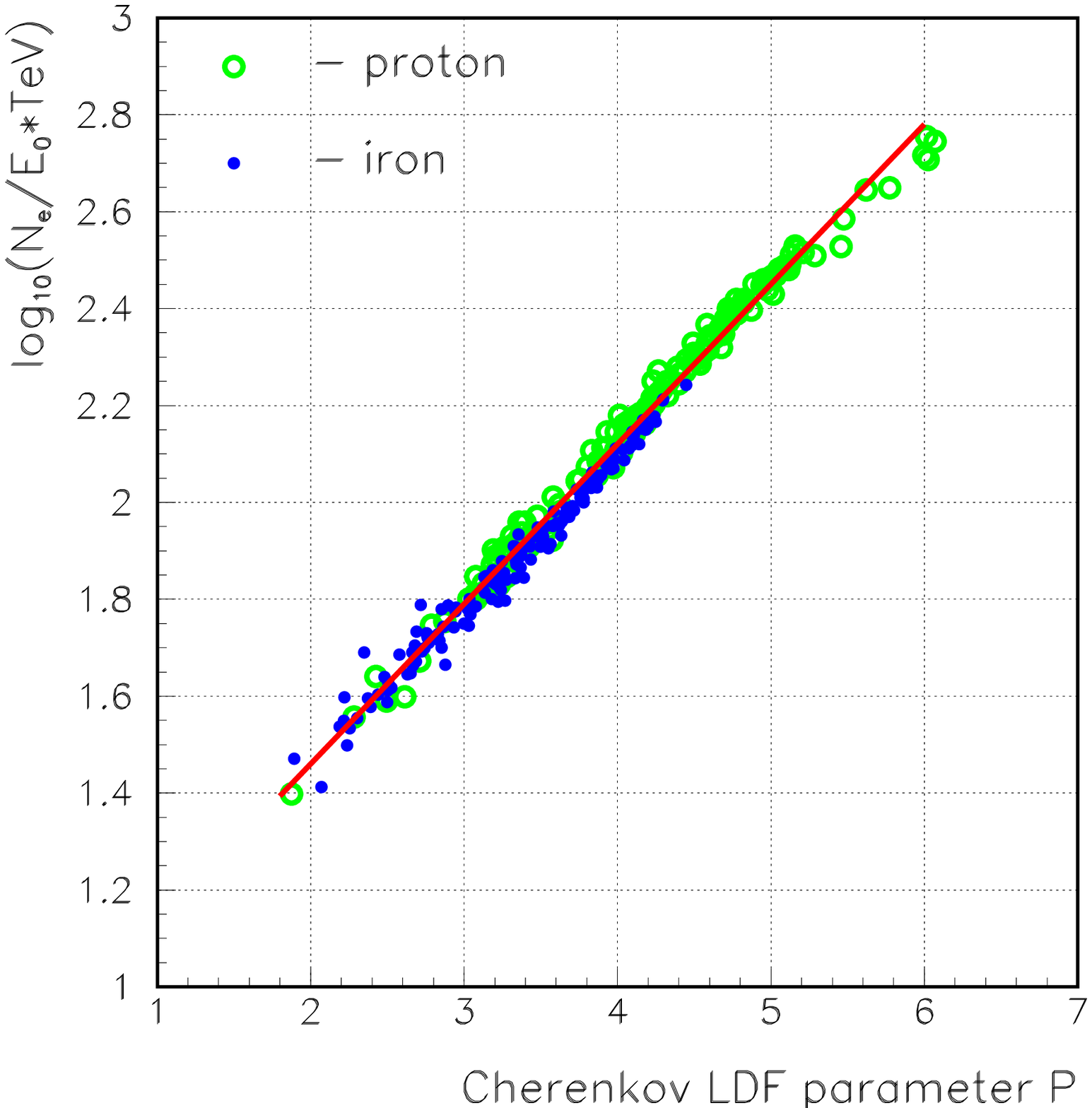,width=0.48\textwidth}
\hfill
 \epsfig{file=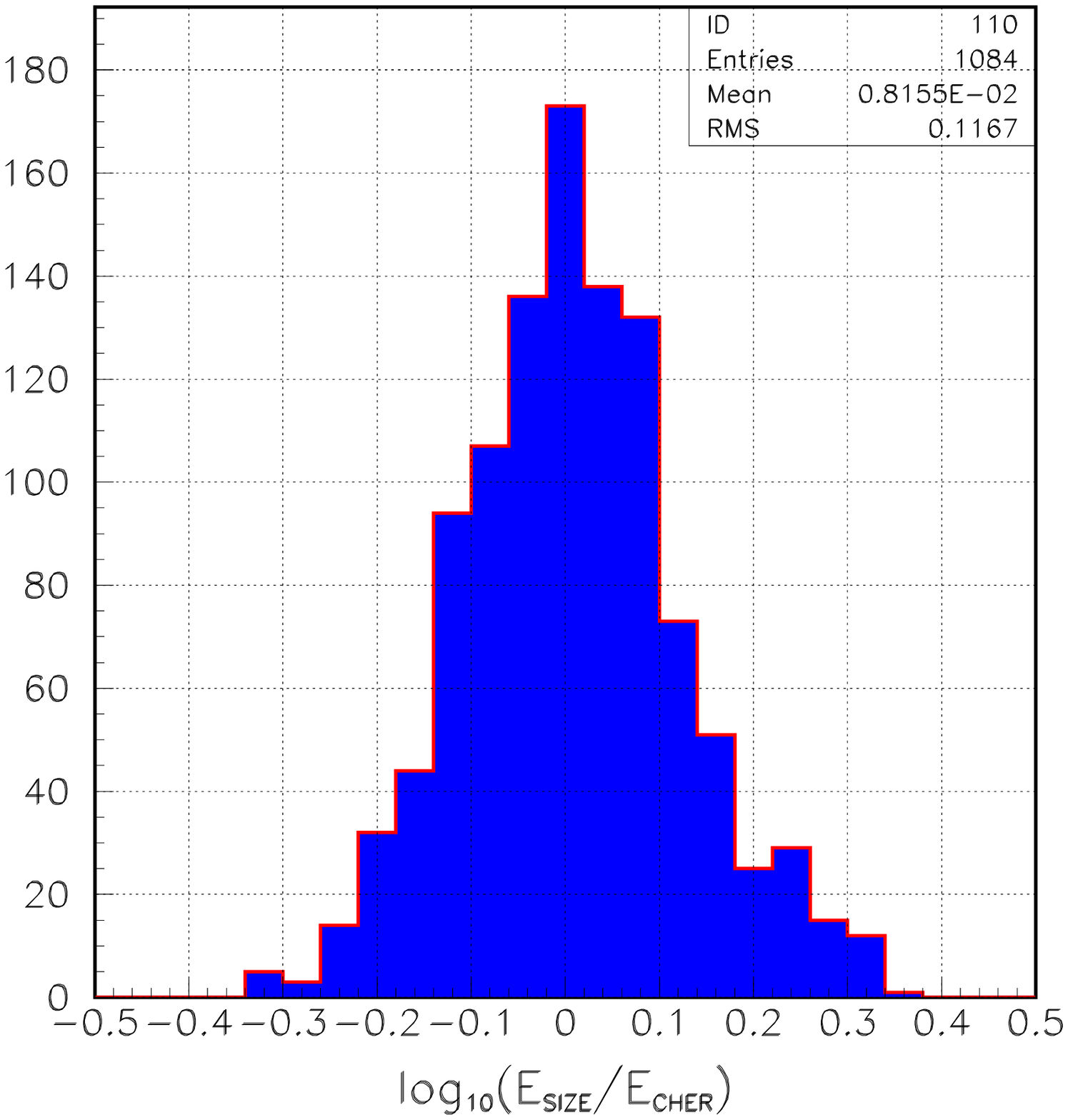,width=0.48\textwidth}
\vspace*{8pt}
\parbox[t]{0.48\textwidth}{\caption{CORSIKA: 
Correlation between EAS Cherenkov LDF steepness and $N_e/E_0$.}}
\hfill
\parbox[t]{0.48\textwidth}{\caption{EXPERIMENT: Comparison of energy, 
obtained with two different methods for $E_{CHER}\geq 2\times10^{15}$~eV.}}
\end{figure}

\section{1st Method: Size and Cherenkov Light LDF steepness P}

To derive the size comparable to the experimental one, we have
taken into account both electrons and muons and used the
experimental procedure of size reconstruction with NKG fitting
function. Figure~1 shows the correlation between the parameter $P$
($P = Q(100)/Q(200)$) and the ratio of the size to primary energy
($N_e/E_0$). Using this relation we can get the primary energy
from measurement of $N_e$ and $P$:
\[
	E_{SIZE}\;[\mbox{eV}] = 1.59\times10^{11}\;N_e/10^{0.33P}. 
\]
This method is quite composition independent, since it gives only $6$\%
of difference in energy estimation for proton and iron. Its main advantage
relies in the well developed technique of scintillator response calibration
based on the measurement of the single particle response\cite{5}. 
From the simulation we derive for the reconstructed energy relative
errors (mostly due to apparatus error of parameter $P$ reconstruction) 
of about $30$\%. 

Similar method of energy reconstruction, but for LDF steepness, estimated at
smaller distances from the core ($20-100$~m), was suggested in Ref.~\refcite{6}.

\section{2nd Method: Cherenkov Light Flux at $175$~m Core Distance $Q_{175}$.}  

From the simulation we obtain the following relation between
$Q_{175}$~[$\mbox{photon}\cdot\mbox{cm}^{-2}\cdot\mbox{eV}^{-1}$]
and primary energy $E_{CHER}$~[eV]: 
\[
	E_{CHER} = 4\times10^{14}\;Q_{175}^{0.94}. 
\]
This method leads to larger differences in
energy estimations for primary proton and iron (about $19$\%), than
the first one. But the error of reconstructed energy (estimated with the above
mentioned simulation of the experiment) is about $15$\% even for the assumed      
``$50$\%~$p$ + $50$\%~Fe'' composition, i.e. two times better than the first one.

The main problem of this method is in the absolute calibration of
Cherenkov light detectors, its error being estimated between $18$\%
to $30$\% for different experiments. To get better accuracy we used
the most probable ratio $E_{SIZE}/E_{CHER}$ as the coefficient for
absolute calibration of Cherenkov detectors response. The coefficient is
estimated for the natural mass composition. After such recalculation, 
the final comparison of two methods in the experimental data is shown in
\figurename~2.
The standard deviation of the experimental distribution is very
close to the one obtained in MC simulation of experiment for the
SIZE/CLDF method (see above).

\section{A Reference Integral Primary Cosmic Rays Intensity}

$600$ events have been recorded during $140$~h of
data taking with energy (derived with the ``re-calibrated Cherenkov
light flux method'') larger than $3\times10^{15}$~eV. The
corresponding integral intensity is: 
\[
I\left(E_0\ge3\times10^{15}\;\mbox{eV}\right)
=(2.3\pm0.1)\times10^{-7} 
\mbox{m}^{-2}\cdot\mbox{s}^{-1}\cdot\mbox{ster}^{-1}.
\]
The statistical uncertainty of the measurement is about $5$\%. 

This estimation can be used as a reference point for other experiments having no
precise absolute calibration, such as Cherenkov light experiment TUNKA\cite{7}. 

\section*{Acknowledgements}

The experimental start of this work has been supported by INTAS grant 96-0526.

\end{document}